\documentstyle[twocolumn,aps,prl,epsfig,subfigure,
amsmath,amsfonts,amssymb]{revtex}
\begin{document}
\twocolumn[\hsize\textwidth\columnwidth\hsize\csname@twocolumnfalse\endcsname
\title{At the very origin of chaos in three-dimensional dynamical systems}
\author{R.~Festa, A.~Mazzino and D.~Vincenzi\\
\small{INFM--Department of Physics, University of Genova, I--16146
Genova, Italy.}}
\draft
\date{\today}
\maketitle
\begin{abstract}
The mechanism responsible for the emergence of chaotic behavior
has been identified analytically within a class of three-dimensional
dynamical systems
which generalize the well-known E.N.~Lorenz 1963  system.
The dynamics in the phase space 
 has been reformulated in terms of a first-exit-time
problem. Chaos emerges due to discontinuous solutions of a
transcendental problem ruling the time for a particle to
cross a potential wall. Numerical results point toward the genericity
of the new found mechanism.
\end{abstract}
]

Many chaotic phenomena in physical sciences 
and engineering can be satisfactorily described by 
low-dimensional autonomous dynamical systems. Among them
the most celebrated example  
is the well-known E.N.~Lorenz model \cite{L63,S82}
\begin{equation}
\left\{ 
\begin{array}{lcl} 
\dot x & = & \sigma ( y - x)\\
\dot y & = &-y + x + (r-1)(1 - z )x\\
\dot z & = & - b ( z - xy)\;, 
\end{array} 
\right. 
\label{ScalLorenz}
\end{equation}
here rewritten using rescaled variables  $x$, $y$, $z$ in the case $r > 1$.
Besides it attracted the attention of the scientific community mainly
on account of its ability to illustrate how a simple model
can produce very rich and varied forms of dynamics, depending 
on the control parameters in the equations. In particular,
chaotic behavior in its 
configuration space around three fixed unstable points
$(0,0,0)$ and $(\pm 1,\pm 1, 1)$ may take place
for $r$ larger than a critical
 value function of $\sigma$ and $b$ \cite{L63,S82}.\\
Starting from the original work of Lorenz \cite{L63}, a 
huge literature 
has grown on the subject (see, e.g., Refs.~\cite{MR99,HUF99,Y00,MS00}
for recent literature), almost 
independently from its somewhat doubtful success of
satisfactorily describing the fluid behavior 
related to the Rayleigh-B\'enard 
problem \cite{S62}. Nevertheless, for the class of Lorenz-like
dynamical systems no determined attempt
has been made yet in order 
to  investigate analitycally 
the very origin of their chaotic behavior. 
This is the main aim of the present work. 
Specifically, we shall reformulate the Lorenz model
dynamics in the phase space
in terms of  more familiar ideas such as mechanical properties
of particles moving in one-dimensional
potential fields subjected to viscous  forcing. Such reformulation
will be the starting point to generalize the original Lorenz system to
a whole class of Lorenz-like systems. 
Among them we shall select a particular piecewise 
linearized model through which
analytical results on the dynamics can be obtained. As we shall see,
the dynamics generated by this model can be mapped into a 
first-exit-time problem. This will allow us to identify
analytically the very  origin of the system chaotic behavior.
Finally, the robustness of the identified mechanism will be tested
numerically for other fully nonlinear generalized Lorenz-like systems.

Let us start our analysis by noticing that, far from the initial transient,
the system (\ref{ScalLorenz}) is equivalent to the 
(integral-) differential equation
\begin{equation}
\ddot x = - (\sigma+1)\dot x - \frac{\partial{\cal U}}{\partial x} .
\label{LorEq}
\end{equation}
Eq.~(\ref{LorEq}) can be interpreted as a customary classical 
dynamical equation of motion for a (unit mass) particle subjected 
to a viscous force $-(\sigma+ 1)\dot x$ 
in the potential field ${\cal U} (x,t)$. Here,
\begin{equation}
{\cal U} (x,t) =  \frac{b}{2\sigma} U(x) + 
\left( 1-\frac{b}{2\sigma}\right)U_t(x)\;
\end{equation}
is a potential field resulting from a weighted average 
(in the chaotic regime $b< 2\sigma$) of a constant quartic potential 
$U(x)\equiv \sigma (r-1)(x^2 - 1)^2/4$ and a time depending quadratic potential
$U_t(x)\equiv \sigma (r-1) [x^2 - 1]_b (x^2 - 1)/2$.\\
We have defined  
\begin{equation}
\left[ f\right]_b \equiv b\int_0^\infty d\tau\;
e^{-b\tau} \; f(t-\tau)\;
\label{memdef}
\end{equation}
where $f(t)$ is a generic function for which the integral exists.
The latter integral clearly represents the memory of  $f(t)$
(at the time $t$), i.e. the present outcome of its exponentially weighted 
past evolution. Thus,
the potential $U_t(x)$  
depends on time through the exponential memory of the past 
motion.\\ 
One can check that, given $x$ from  (\ref{LorEq}), the
variables $y$ and $z$ are obtained from the relations $y=\dot x/\sigma
+ x$
and $z=b/(2\sigma)x^2+(1-b/(2\sigma))[x^2]_b$, respectively. 
In this formulation it is for instance evident that $x$ and $y$  are
synchronazing coordinates  while this is not for $z$ \cite{MR99,PCJM97}.

The above description immediately leads to the following considerations
highlighting
the role of the memory in the route to chaos. For $U_t=0$, 
the particle motion
stops after some time in one of the two minima of $U(x)$. This because of the viscous
term $(\sigma +1)\dot x$. On the
contrary, non trivial behaviors, including chaotic ones, may take
place due to the statistical balance between dissipation and energy
exchanges produced by $U_t$. The bistable character of $U$ plays a crucial role
for the emergence of chaos. Indeed, initially very close 
trajectories starting in the same  cell, may undergo to completely
different evolutions if, at a certain time, one has sufficiently energy
to cross the peak in $x=0$, while this is not for the other. 
This is the first clue that chaos arises from the unpredictability
of the instants when particles pass through the barrier in $x=0$.
We shall give in the sequel the analytical proof of this heuristic 
argument together with the reason at the 
very origin of such unpredictability.
To do that, let us give a further reformulation of the system 
(\ref{ScalLorenz}). This will make it possible 
to generalize and simplify (\ref{ScalLorenz}) with the final goal
to deal with chaos analytically.\\
By suitably scaling the time: 
$t\mapsto \tau \equiv [{(r-1)/2}]^{1/2}\;t$, Eq.~(\ref{LorEq}) can
be recast in the form \cite{duff}
\begin{equation}
\ddot x + \eta \dot x 
+ (x^2 - 1)\; x = -\alpha \left[x^2 -1\right]_\beta\;x
\label{LorEqSt}
\end{equation}
where $\eta$, $\alpha$ and $\beta$ are related to the original
parameters $\sigma$, $b$ and $r$.\\
A more expressive form of Eq.~(\ref{LorEqSt}) is
\begin{equation}
\ddot x + \eta \dot x 
+ \left( q(x) +\alpha 
\left[q(x)\right]_\beta\right)\Phi'(x)= 0\;,
\label{LorEqSt1}
\end{equation}
where $ \Phi(x)= 1/2\; x^2$ and $q(x) = x^2- 1 =2 \Phi (x) -1$.
This equation describes
the motion of a (unit mass) particle subjected to a viscous force
$-\eta \dot x$ and interacting with  a fixed potential field 
$\Phi(x)$ through a `dynamically varying charge' $
q_t(x) = q(x) + \alpha \left[q(x)\right]_\beta $. It is
constituted by a fixed `core' charge, a 
`locally acquired' charge, related to the local
potential, and an exponentially vanishing `memory' charge, 
continuously depending on the previous instantaneous charge 
history. This scheme can be used to mimic the instantaneous effective
charge of a particle moving in a background (structured) particle
bath. Indeed, the coupling of $\left[q(x)\right]_\beta$ with the 
background potential  $\Phi(x)$ yields an endogenous  forcing 
term which allows a self-sustained unceasing motion, even in the 
presence of friction, and causes a corresponding unceasing inner
transfer of the bath charge .

At this point the Lorenz system (\ref{ScalLorenz}) can be easily
generalized in the form
\begin{equation}
\left\{ 
\begin{array}{lcl} 
\dot x & = & \sigma ( y - x)\\
\dot y & = &-y + x + (r-1)(1 - z )\Phi'(x)\\
\dot z & = & - b \left[ z - \frac{1}{2}q'(x) (y-x) - q(x) - 
1\right]\;, 
\end{array} \right.
\label{GenLorenz}
\end{equation} 
from which Eq.~(\ref{LorEqSt1}) follows after some algebraic manipulations.

As already observed,
the chaotic behavior of Lorenz's system essentially depends on 
the unpredictability of the instants when $x$ 
change its sign: as long as it keeps constant sign the system 
evolution is certainly nonlinear, and nevertheless not chaotic 
at all. This fact suggests a slight modification of the original 
form, in order to single out analytically the origin of chaos without to be 
faced with the difficulties arising from nonlinear 
problems. We thus set in Eqs.~(\ref{LorEqSt1}) and (\ref{GenLorenz}) 
$\Phi(x)= |x|$ and $q(x)= \Phi (x) - 1$
obtaining:
\begin{equation}
\ddot x + \eta\dot x + \left\{|x|- 1+ \alpha\left[|x|-
1\right]_b\right\}\; 
{\rm sgn}(x)=0
\label{LinLorEq1}
\end{equation}
and the corresponding piecewise linearized system
\begin{equation}
\left\{ 
\begin{array}{lcl} 
\dot x & = & \sigma ( y - x)\\
\dot y & = &-y + x + (r-1)(1 - z )\; {\rm sgn}(x)\\
\dot z & = & - b z + b \; {\rm sgn}(x) \frac{x+y}{2}\;
\end{array} \right.
\label{LinLorenz}
\end{equation} 
where ${\rm sgn}(x)\equiv|x|/x$.\\
Our assumptions for $\Phi$ and $q$ correspond in Eq.~(\ref{LorEq})
to 
$U(x)=\sigma(r-1)(|x|-1)^2/2$ and $U_t(x)=\sigma(r-1)[|x|-1]_{\beta}
(|x|-1)$. The fundamental bistable character of $U$ is thus 
maintained.\\
%
In order to solve  Eq.~(\ref{LinLorEq1}), we exploit the fact that 
it is left invariant under the transformations
$x\mapsto - x$, $\dot x\mapsto - \dot x$ and $\ddot x
\mapsto -\ddot x$. We can thus focus on one of the two regions  
$x<0$ and $x>0$. Let us consider, e.g., $x>0$ .
In this case Eq.~(\ref{LinLorEq1}) is 
equivalent to
the third order linear equation
\begin{equation}
\frac{d^3 \xi}{d\tau^3}+ (\beta+ \eta)\frac{d^2 \xi}{d\tau^2} 
+ (1 + \beta \eta)\frac{d \xi}{d\tau}+\beta (1+\alpha) \xi = 0
\label{Third}
\end{equation}
where $\xi \equiv x - 1$.\\
Notice that when the particles cross the barrier in $x=0$, 
the evolution described by Eq.~(\ref{Third}) has to restart with 
new initial conditions corresponding to an elastic collision against 
a rigid wall posed in $x=0$.\\
It can be checked that, in the case of chaotic motion we are 
interested in, the general solution of Eq.~(\ref{Third})
can always be written  in the form
\begin{equation}
\xi(\tau)=  e^{\lambda_r \tau}\left( C_1 \cos (\lambda_i \tau) + 
C_2 \sin (\lambda_i \tau)\right) + C_3e^{-\lambda_0 \tau}\;
\label{Solution}
\end{equation}
with $\lambda_0,\lambda_r,\lambda_i \ge 0$ and $C_1$, $C_2$ and $C_3$ 
real coefficients determined from the initial conditions on 
$\xi$, $\dot\xi$  and $\ddot\xi$.\\
The instant $\tau_1$ at which the first particle collision against the
wall in $x=0$ occurs is thus defined by the equation $\xi (\tau_1)=-1$.
 From (\ref{Solution}), $\tau_1$ is thus the smallest positive
solution of the transcendental equation
\begin{equation}
\label{trasc.}
C_1\, e^{\lambda_r \tau_1}\cos (\lambda_i \tau_1)\, +\,
   C_2 \,e^{\lambda_r \tau_1}\sin (\lambda_i \tau_1)\, +\,
   C_3\, e^{-\lambda_0 \tau_1}=-1. 
\end{equation}
Geometrically, we can interpret
$\tau_1$ as the first intersection of 
$g(\tau)\equiv -C_3\, e^{-\lambda_0 \tau} -1$ 
with $h(\tau)\equiv 
C_1\, e^{\lambda_r \tau}\cos (\lambda_i \tau)\, +\,
   C_2\, e^{\lambda_r \tau}\sin (\lambda_i \tau)$.
The function $g$ is a decreasing exponential and
$h$ an oscillating function with growing amplitude.
It is thus easily understood why  even a little 
modification of initial conditions can produce 
a discontinuous variation of $\tau_1$ (see Fig.~1). 
As we shall see, the same reason applies also for the class
of systems (\ref{GenLorenz}). 
\begin{figure}[!h]
\epsfxsize=8.5truecm
\vspace{-1cm}
\epsfbox{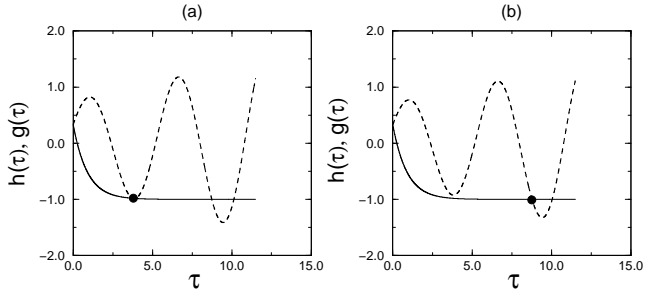}
\vspace{-2cm}
\caption{Graphical interpretation of the discontinuous character of 
$\tau_1$ for small changes of the initial conditions. Full line 
represents $g(\tau)$, dashed lines $h(\tau)$
for the parameters $\alpha =6.50$, $\beta=0.19$, $\eta=0.78$ and
for initial conditions (a): $\dot{\xi}_0=2.26$, $\ddot{\xi}_0=-2.00$;
(b): $\dot{\xi}_0=2.30$, $\ddot{\xi}_0=-2.00$.
Bullets denote the first intersection between $g$ and $h$ defining
the first collision time against the wall
(a): $\tau_1 = 3.91$; (b) $\tau_1 =8.76$.}
\label{fig1}
\end{figure}
\noindent The dependence of $\tau_1$ 
on the initial conditions $\dot{\xi}_0$
and $\ddot\xi_0$ ($\xi_0=-1$) is implicitly contained 
in Eq.~(\ref{trasc.}). Coefficients $C_1$, $C_2$ and $C_3$ are
indeed linearly related to the initial conditions. The way to describe
the global behavior of $\tau_1$ in terms of the pair 
$(\dot{\xi}_0,\ddot{\xi}_0)$
although simple
results quite lengthy. We thus confine our attention on the
corresponding graphical shape presented in Fig.~2.
\begin{figure}[!h]
\epsfxsize=8.5truecm
\epsfbox{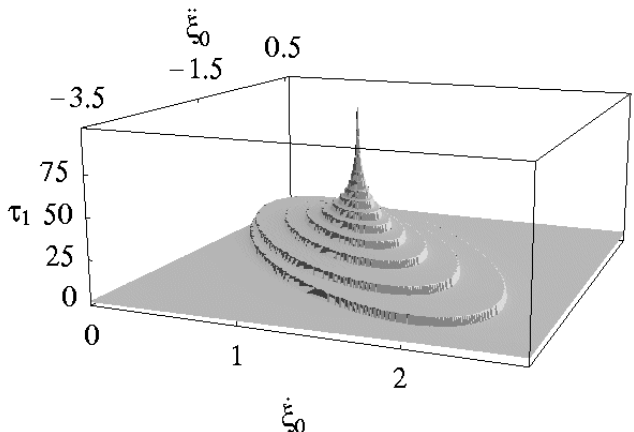}
\caption{The behavior of the first collision time, $\tau_1$ defined
by Eq.~(\protect\ref{trasc.}), {\it vs} the initial conditions  $\dot{\xi}_0$
and $\ddot\xi_0$
for $\alpha =6.50$, $\beta=0.19$, $\eta=0.78$.}
\label{fig2}
\end{figure}
\noindent From the figure it appears that $\tau_1$ shows sensitivity
with respect to the initial conditions only in a 
limited subset of the half-plane $\dot{\xi}_0\geq 0$.
It is natural to expect  that in the chaotic regime the system
is quickly attracted inside this region.
It is now clear why chaotic behaviors can arise also for 
apparently simple nonlinearities as isolated (non eliminable)
discontinuities (see, e.g., Refs.~\cite{Andronov,Sparrow1,Holmes,Chua,L78}
for other examples of 
piecewise linearized systems displaying varied form of chaotic behaviors).\\
Notice that the same behavior showed in Fig.~2
for $\tau_1$ holds for the $n$-th
collision time $\tau_n$ as a function of the $(n-1)$-th initial
conditions $(\dot{\xi}_0^{(n-1)}, \ddot{\xi}_0^{(n-1)})$. \\
For $(\dot{\xi}_0,
\ddot{\xi}_0)= (\lambda_0,-\lambda_0^2)$ 
one has $\tau_1\to\infty$ (the peak in Fig.~2). 
In this case, it is possible to show that
the system configuration, described by 
$ \boldsymbol{\xi}\equiv (\xi (\tau), \dot{\xi}(\tau),
\ddot{\xi}(\tau))$, 
exactly lies on the stable manifold
${\mathcal W}^s\equiv\{t\;\boldsymbol{w}_3|\;t>-1\}$ with 
$\boldsymbol{w}_3\equiv (1,-\lambda_0,\lambda_0^2)$ 
and its motion is an exponential 
decay on the fixed point $\boldsymbol{\xi}=\boldsymbol{0}$. \\
 For $(\dot{\xi}_0,
\ddot{\xi}_0)\neq (\lambda_0,-\lambda_0^2)$
the evolution in the phase space
$\xi$, $\dot{\xi}$, $\ddot{\xi}$ consists
of both a `rapid' decay towards $\boldsymbol{\xi}=
\boldsymbol{0}$ along the stable manifold 
${\mathcal W}^s$ and a `slow' amplified
oscillation on the two-dimensional unstable manifold, 
${\mathcal W}^u$, generated by 
$\boldsymbol{w}_1\equiv (1,\lambda_c, \lambda_c^2)$ and
$\boldsymbol{w}_2\equiv (1,\bar{\lambda}_c, \bar{\lambda}_c^2)$,
with $\lambda_c=\lambda_r+i\lambda_i$ and 
$\bar{\lambda}_c=\lambda_r-i\lambda_i$.\\
Focusing now on the $n$-th collision, we thus have from Eq.~(\ref{trasc.})
the behavior of $\tau_n$ {\it vs}
the pair  $(\dot{\xi}_0^{(n-1)},\ddot{\xi}_0^{(n-1)})$.
Taking the first and the second time derivative of
Eq.~(\ref{trasc.}),  we can relate $\dot{\xi}_0^{(n)}$ and 
$\ddot{\xi}_0^{(n)}$ to the initial conditions $\dot{\xi}_0^{(n-1)}$ and 
$\ddot{\xi}_0^{(n-1)}$ (remember that $C_1$, $C_2$ and $ C_3$ are
linearly related to such initial conditions). 
The result is
a two-dimensional Poincar\'e map connecting 
$(\dot{\xi}_0^{(n-1)},\ddot{\xi}_0^{(n-1)})$  to 
$(\dot{\xi}_0^{(n)},\ddot{\xi}_0^{(n)})$.
The nonlinear character of the map is entirely contained in the
transcendental and discontinuos 
relation between $\tau_n$ and the initial conditions.\\
\begin{figure}[!h]
\epsfxsize=8.5truecm
\vspace{-1cm}
\epsfbox{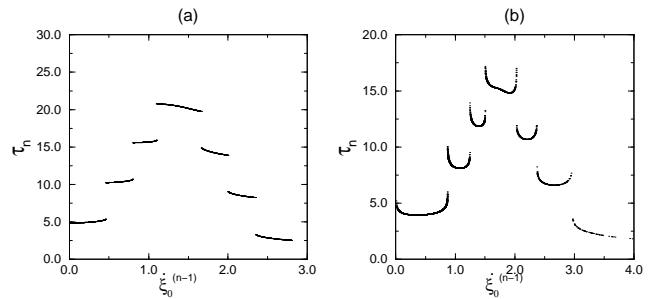}
\vspace{-2cm}
\caption{The one-dimensional map
$\tau_n=\tau_n(\dot{\xi}_0^{(n-1)})$; (a):
for the piecewise linearized system (\ref{LinLorenz});
(b):
for the Lorenz model (\ref{ScalLorenz}).}
\label{fig3}
\end{figure}
Let us now assume that the attraction towards the unstable manifold
${\mathcal W}^u$ is sufficiently fast (the goodness of this approximation
is controlled by $\beta+\eta$) for the system to be considered 
as belonging to ${\mathcal W}^u$ at the collision time 
against the plane $\xi=-1$. If this is the case, `immediately'
after each collision the system thus necessary lies on a
straight line. 
This implies a  linear relationship between velocity 
and acceleration immediately after the $n$-th collision.
The functional dependence 
of $\tau_n$ on $\dot{\xi}_0^{(n-1)}$ and $\ddot{\xi}_0^{(n-1)}$ is thus
restricted on a straight line in the plane 
$\dot{\xi}_0^{(n-1)}$ -- $\ddot{\xi}_0^{(n-1)}$. This amounts to say
that $\tau_n$ is a function of one variable alone, say, 
$\dot{\xi}_0^{(n-1)}$. The map $\tau_n=\tau_n(\dot{\xi}_0^{(n-1)})$
is shown in Fig.~3(a).\\
\noindent Furthermore,
the Poincar\'e map connecting $(\dot{\xi}_0^{(n-1)},\ddot{\xi}_0^{(n-1)})$
to $(\dot{\xi}_0^{(n)},\ddot{\xi}_0^{(n)})$ reduces to a
one-dimensional map, say, between 
$\dot{\xi}_0^{(n-1)}$ and $\dot{\xi}_0^{(n)}$. Fig.~4(a) shows
this map for
$\alpha =6.50$, $\beta=0.19$, $\eta=0.78$.
Its behavior is evidently chaotic: the discontinuities
 derive from the analogous ones
in the function  $\tau_n= \tau_n(\dot{\xi}_0^{(n-1)})$.\\
\begin{figure}[!h]
\epsfxsize=8.5truecm
\vspace{-1cm}
\epsfbox{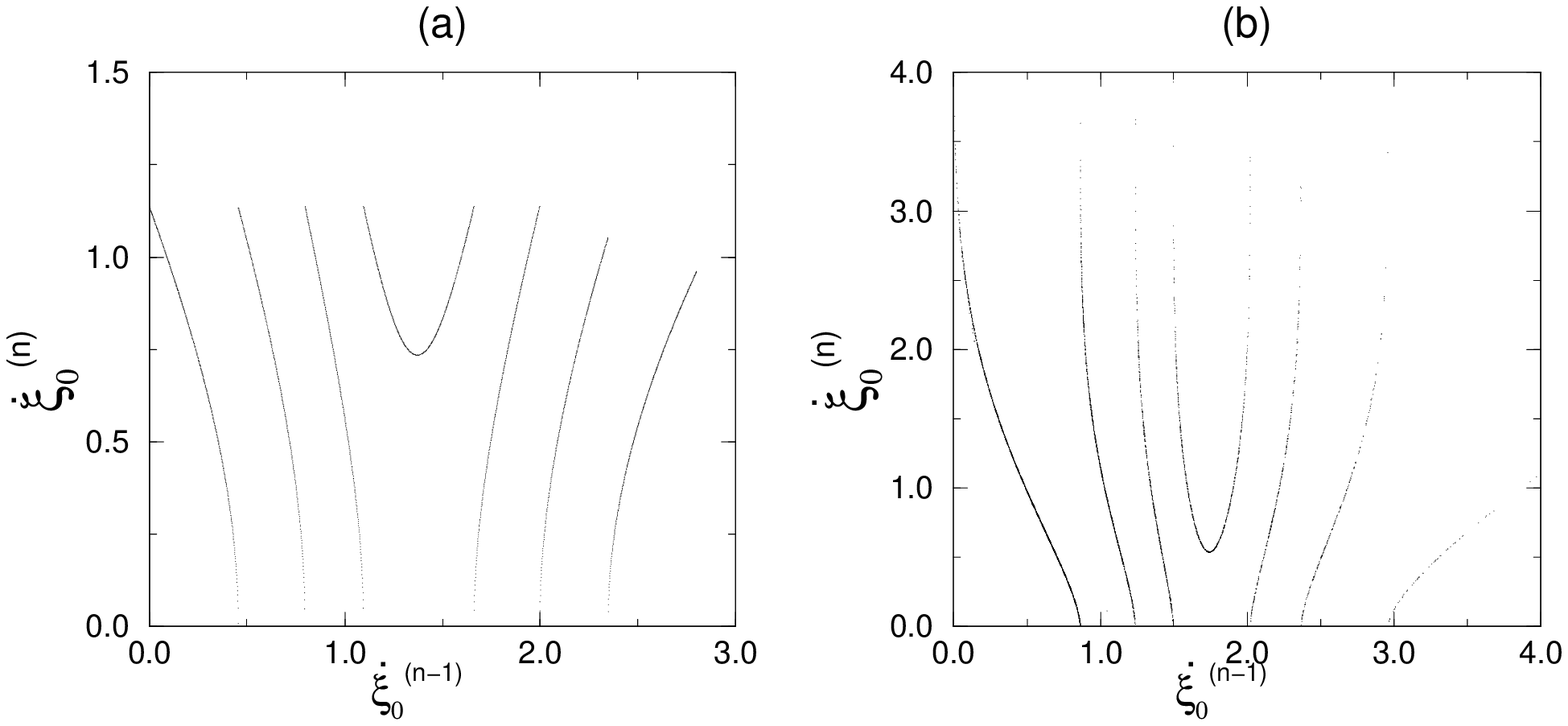}
\vspace{-2cm}
\caption{The one-dimensional map between
$\dot{\xi}_0^{(n)}$ and $\dot{\xi}_0^{(n-1)}$; (a):
for the piecewise linearized system (\ref{LinLorenz});
(b): for the Lorenz model (\ref{ScalLorenz}).}
\label{fig4}
\end{figure}
The final question to be addressed concerns the 
relation of our results with the original 
Lorenz system (\ref{ScalLorenz}) and more generally with the class
of dynamical systems (\ref{GenLorenz}). One immediately realizes that,
under the condition $\Phi^{'}(0)=0$, the linear relationship between
velocity and acceleration in $x=0$ follows:
\begin{equation}
\ddot{x}_0^{(n)}+\eta \dot{x}_0^{(n)}=0 .
\end{equation}
As for the piecewise linearized system (\ref{LinLorenz}), the
generalized Lorenz systems  (\ref{GenLorenz}) are thus described 
by one-dimensional maps. Focusing in particular on the
Lorenz model (\ref{ScalLorenz}), the analogous of the maps reported
in Figs.~3(a) and 4(a) are shown in Figs.~3(b) and 4(b). Similar
behaviors have been obtained for other choices of $\Phi$ and $q$ 
in (\ref{GenLorenz}). These maps have been derived by
numerical integration of (\ref{GenLorenz}) by a standard Runge--Kutta
scheme, whereas we remember that all results relative to the 
piecewise linearized system have been obtained analytically.\\
The resemblance of Figs.~3(a), 4(a) with 
Figs.~3(b), 4(b) points toward the robustness of the mechanism
we have identified as cause of chaos in the linearized 
system (\ref{LinLorenz}).

In conclusion, the very origin of chaos for a whole class of
three-dimensional autonomous dynamical systems has been identified.
Chaos is entirely contained in a transcendental equation
ruling a first-exit-time problem 
whose solutions appear discontinuous for
small changes in the
initial conditions. Results have been obtained analytically for a
piecewise linearized model belonging to a more general class
of dynamical systems. We however showed numerically that the basic reason 
for the chaos to emerge applies also for the general case.

We thank G.~Cassinelli, A.~Celani, M.~La Camera, M.~Vergassola
and A.~Vulpiani for illuminating discussions and suggestions.
AM has been partially supported by the INFM project GEPAIGG01.


\begin{thebibliography}{99}
\bibitem{duff}
Eq.~(\ref{LorEqSt}) resembles other examples of chaotic nonlinear 
nonautonomous equations, as, for instance, the inverse 
Duffing equation $
\ddot x + \eta \dot x 
+ (x^2 - 1)\; x = A\cos (\Omega t)\;,$
which describes a  sinusoidally forced anharmonic oscillator: in 
our case the motion is  self-sustained through the generally non 
sinusoidal `autonomous' forcing term $-\alpha \left[x^2 
-1\right]_\beta\;x$.
\end{thebibliography}
\end{document}